\documentclass[aps,showpacs,preprintnumbers,amsmath,amssymb]{revtex4}

\usepackage{graphics,epsfig}
\usepackage{graphicx}
\usepackage{dcolumn}

\newcommand{\be}{\begin{equation}}
\newcommand{\ee}{\end{equation}}
\newcommand{\ba}{\begin{eqnarray}}
\newcommand{\ea}{\end{eqnarray}}

\begin{document}
\title{{\bf Quasinormal modes of a
 black hole surrounded by quintessence}}
\author{Songbai Chen}\email{csb3752@hotmail.com}
\author{Jiliang Jing} \email{jljing@hunnu.edu.cn}
\affiliation{ Institute of Physics and  Department of Physics, \\
Hunan Normal University,  Changsha, Hunan 410081, P. R. China }

\vspace*{0.2cm}
\begin{abstract}

Using the third-order WKB approximation, we evaluate the
quasinormal frequencies of massless scalar field perturbation
around the black hole which is surrounded by the static and
spherically symmetric quintessence.  Our result shows that due to
the presence of quintessence, the scalar field damps more rapidly.
Moreover, we also note that the quintessential state parameter
$\epsilon$ (the ratio of pressure $p_q$ to the energy density
$\rho_q$) play an important role for the quasinormal frequencies.
As the state parameter $\epsilon$ increases the real part
increases and the absolute value of the imaginary part decreases.
This means that the scalar field decays more slowly in the larger
$\epsilon$ quintessence case.
\end{abstract}

 \maketitle

%\newpage
%\vspace*{0.2cm}
\section{Introduction}

The accelerated expansion is one of the most important aspects of
the present universe. It seems to be strongly supported by the
recent cosmological observations, such as supernova of the type Ia
\cite{AAS}, the anisotropy of the cosmic microwave background
radiation \cite{PAS}, large scale structure \cite{RMT}, and so on.
However, the nature of leading to the cosmic accelerated expansion
is still uncertain. In terms of the Einstein's gravity theory,
this accelerated expansion can be explained by the conjecture that
at late times the considerable part of the total density of the
universe is dominated by dark energy with a negative pressure
\cite{PBTT}. There are several candidates for the dark energy: the
vacuum energy (the cosmological constant) \cite{Swei} and
dynamical components (such as quintessence \cite{BRa}, k-essence
\cite{CAR} and phantom models \cite{RRCa}\cite{BMcI}). Models of
dark energy differ with respect to the size of the parameter
$\epsilon$, namely, the relation between the pressure and energy
density of dark energy. For example, in the cosmological constant
model, $\epsilon$ maintains the constant $-1$. For the
quintessence, $\epsilon$ keeps in the range of $-1\leq\epsilon
<0$. Obviously, the cosmological constant can be treated as a
special quintessence. Recently, these theoretical models have
attracted much attention because that they possess many advantages
in the explanation of the cosmological accelerated expansion.
However, whether the dark energy is real and unique reason of this
accelerated expansion need to be further verified in the future.

On the other hand, black hole is another fascinating object in the
modern physics. Since Hawking radiation \cite{Hawking} was
discovered, a great deal attention has been focused on many fields
in the black hole physics including the black hole entropy
\cite{JDB}-\cite{Hooft}, loss of information paradox
\cite{SHa}-\cite{SHod1}, holographic
principle\cite{JDB1}-\cite{LSU}, and so on. It is believed widely
that the study of black hole may lead to a deeper understanding of
the relationship among the general relative theory, quantum
mechanics, thermodynamics and statistics. This means that black
hole physics will play an important role in the fundamental
physics. However, at present whether black hole exists in our
universe or not is still unclear. Recent investigation shows that
quasinormal modes can provide a direct way to identify a black
hole existence in the universe because that they carry the
characteristic information of black holes
\cite{Chandrasekhar}\cite{Regge}. Moreover, it is also found that
the quasinormal modes have a close connection with the AdS/CFT
correspondence \cite{Witten}\cite{Maldacena}\cite{Kalyana} and the
Loop Quantum Gravity \cite{Hod98}\cite{Dreyer03}. Thus, the study
of quasinormal modes in black hole spacetimes becomes appealing
recent years \cite{Cardoso}-\cite{Jing2}.

From the above discussion, we know that the dark energy and the
quasinormal modes are two hot topics in the physics at present.
Then it is naturally to raise the question whether the dark energy
affects the quasinormal modes.  From the literatures
\cite{Zhidenko}\cite{Moss}, we know that due to presence of the
cosmological constant, quasinormal frequencies of a black hole in
de Sitter spacetime are clearly different from that in the
asymptotically flat metrics. It seems to imply that there exist
some certain connections between dark energy and quasinormal modes
because that the cosmological constant can be regarded as a model
of dark energy in the cosmology. However, how the dark energy
affects the quasinormal modes remains unclear in my best
knowledge. Recently, Kiselev \cite{VVK} considered Einstein's
field equations for a black hole surrounded by the static and
spherically symmetric quintessence whose energy-momentum tensor
satisfied the additive and linear conditions, and obtained a new
static solution which depends on the state parameter $\epsilon$ of
the quintessence. It is very similar to the black hole solution in
the de Sitter/anti de Sitter spacetime. In this paper,  as a
concrete case, we try to probe the relationship between the
quasinormal modes and dark energy through the evaluation of the
quasinormal modes of massless scalar field perturbation around the
black hole which is surrounded by the quintessence.

The plan of the paper is as follows. In Sec.2 we review briefly
the Kiselev's work and present the metric for a black hole
surrounded by the static and spherically symmetric quintessence.
In Sec.3 we evaluate the quasinormal modes of the massless scalar
field in this spacetime by using the third-order WKB approximation
\cite{BC}\cite{SC}\cite{SI}. At last, we present a summary and
conclusion.

 \vspace*{0.2cm}
\section{The metric for a black hole surrounded by quintessence }
In this section, let us review Kiselev's work \cite{VVK} briefly.
The general metric for the static and spherically symmetric
spacetime is
\begin{eqnarray} \label{metric0}
ds^2=e^{\nu}dt^2-e^{\lambda}dr^2-r^2(d\theta^2+\sin{\theta}^2d\phi^2),
\end{eqnarray}
where $\nu$ and $\lambda$ are functions of $r$. The Einstein field
equations can be written as
\begin{eqnarray} \label{EFeq}
2T^{\;t}_t&=&-e^{-\lambda}\bigg(\frac{1}{r^2}
-\frac{\lambda'}{r}\bigg)+\frac{1}{r^2},\\
2T^{\;r}_r&=&-e^{-\lambda}\bigg(\frac{1}{r^2}
-\frac{\nu'}{r}\bigg)+\frac{1}{r^2},\\
2T^{\;\theta}_{\theta}&=&2T^{\;\phi}_{\phi}=-\frac{1}{2}e^{-\lambda}\bigg(\nu^{''}+
\frac{\nu'}{2}-\frac{\nu'-\lambda'}{r}-\frac{\nu'\lambda'}{2}\bigg),
\end{eqnarray}
where the prime denotes the derivative with respect to $r$. The
energy-momentum tensor of the quintessence which satisfies the
condition of the additivity and linearity can be expressed as
\begin{eqnarray} \label{tensor}
&&T^{\;t}_t=T^{\;r}_r=\rho_q,\nonumber\\
&&T^{\;\theta}_{\theta}=T^{\;\phi}_{\theta}=-\frac{1}{2}\rho_q(3\epsilon+1),
\end{eqnarray}
where the parameter $\epsilon$ is the state parameter of
quintessence. Setting $\lambda=-\ln{(1+f)}$,  we find that the
variable $f$ satisfies the equation
\begin{eqnarray} \label{feq}
r^2f^{''}+3(\epsilon+1)r f'+(3\epsilon+1)f=0.
\end{eqnarray}
This second-order differential equation can be solved exactly and
its general solution is given by
\begin{eqnarray} \label{f}
f=1-\frac{r_g}{r}-\frac{c}{r^{3\epsilon+1}},
\end{eqnarray}
where c and $r_g$ are the normalization factors. Thus, the general
forms of exact spherically symmetric solutions for the Einstein
equations describing black holes surrounded by the quintessential
matter with the energy-momentum tensor, which satisfies the
condition of the additivity and linearity, can be expressed by
\begin{eqnarray} \label{metric1}
ds^2= \bigg(1-\frac{2M}{r}-\frac{c}{r^{3\epsilon+1}}\bigg)dt^2
-\bigg(1-\frac{2M}{r}-\frac{c}{r^{3\epsilon+1}} \bigg)^{-1}dr^2
-r^2(d\theta^2+\sin{\theta}^2d\phi^2),
\end{eqnarray}
where $M$ is the black hole mass. Obviously, it is very similar to
the de Sitter/anti-de Sitter metric. Moreover, we also note that
the well-known Schwarzschild and Reissner-Nordstr\"{o}m metrics
are its special cases. It is not surprising because that the
Schwarzschild and Reissner-Nordstr\"{o}m metrics describe the
black holes surrounded by vacuum and electric fields respectively.

\section{quasinormal mode of a scalar field in the
 black hole surrounded by quintessence}

The general perturbation equation for the massless scalar field in
the curve spacetime is given by
 \begin{eqnarray}
 \frac{1}{\sqrt{-g}}\partial_\mu(\sqrt{-g}g^{\mu\nu}
 \partial_\nu)\psi=0,\label{eq1}
 \end{eqnarray}
where $\psi$ is the scalar field.

Introducing the variables $\psi=\frac{e^{-i\omega
t}\Phi(r)}{r}Y(\theta,\varphi)$ and  $r_*=
\int{\bigg(1-\frac{2M}{r}-\frac{c}{r^{3\epsilon+1}}
\bigg)^{-1}dr}$, and substituting Eq.(\ref{metric1}) into
Eq.(\ref{eq1}), we obtain a radial perturbation equation
\begin{eqnarray}
\frac{d^2\Phi(r)}{dr_*^2}+(\omega^2-V)\Phi(r)=0,\label{e3}
\end{eqnarray}
where
\begin{eqnarray}
V= \left(1-\frac{2M}{r}-\frac{c}{r^{3\epsilon+1}}\right)\left(
\frac{l(l+1)}{r^2}+\frac{2M}{r^3}+\frac{c(3\epsilon+1)}{r^{3\epsilon+3}}
\right)\label{v}.
 \end{eqnarray}

It is obvious that the effective potential $V$ depends only on the
value of $r$, angular quantum number $l$ and ratio $\epsilon$.
Fig. 1 shows the variation of the effective potential with respect
to the state parameter $\epsilon$ of quintessence for fixed $l=5$
and $c=0.001$. From this figure we can find that as the absolute
value $\epsilon$ increases, the peak value of potential barrier
gets lower and the location of the peak ($r=r_p \;$) moves along
the right.
\begin{figure}[ht]
\begin{center}
\includegraphics[width=9cm]{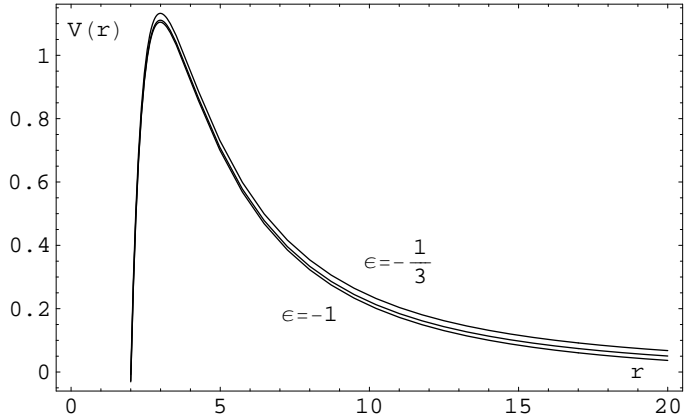}
\caption{ Variation of the effective potential for the massless
scalar field in the black hole surrounded by quintessence with $r$
for fixed $l=5$, $c=0.001$ and $ \epsilon = -1/3$, $-0.9$, $-1$.}
\end{center}
\label{fig1}
\end{figure}

From Eq.(\ref{v}) and Fig.1, we find that the quasinormal
frequencies depends the normalize factor $c$ and state parameter
$\epsilon$ of quintessence. However, in this paper, we only want
to study the relationship between the quasinormal mode and state
parameter $\epsilon$. Thereafter, we take $M=1$ and $c=0.001$ in
our calculation. Let us now evaluate the quasinormal frequencies
for the massless scalar field in the background spacetime
(\ref{metric1}) by using the third-order WKB approximation, a
numerical method devised by Schutz, Will and Iyer
\cite{BC}\cite{SC}\cite{SI}. This method has been used extensively
in evaluating quasinormal frequencies of various black holes
because of its considerable accuracy for lower-lying modes. In
this approximate method, the formula for the complex quasinormal
frequencies $\omega$ is
\begin{eqnarray}
\omega^2=[V_0+(-2V^{''}_0)^{1/2}\Lambda]-i(n+\frac{1}{2})(-2V^{''}_0)^{1/2}(1+\Omega),
\end{eqnarray}
where
\begin{eqnarray}
\Lambda&=&\frac{1}{(-2V^{''}_0)^{1/2}}\left\{\frac{1}{8}\left(\frac{V^{(4)}_0}{V^{''}_0}\right)
\left(\frac{1}{4}+\alpha^2\right)-\frac{1}{288}\left(\frac{V^{'''}_0}{V^{''}_0}\right)^2
(7+60\alpha^2)\right\},\nonumber\\
\Omega&=&\frac{1}{(-2V^{''}_0)^{1/2}}\bigg\{\frac{5}{6912}
\left(\frac{V^{'''}_0}{V^{''}_0}\right)^4
(77+188\alpha^2)\nonumber\\&-&
\frac{1}{384}\left(\frac{V^{'''^2}_0V^{(4)}_0}{V^{''^3}_0}\right)
(51+100\alpha^2)
+\frac{1}{2304}\left(\frac{V^{(4)}_0}{V^{''}_0}\right)^2(67+68\alpha^2)
\nonumber\\&+&\frac{1}{288}
\left(\frac{V^{'''}_0V^{(5)}_0}{V^{''^2}_0}\right)(19+28\alpha^2)-\frac{1}{288}
\left(\frac{V^{(6)}_0}{V^{''}_0}\right)(5+4\alpha^2)\bigg\},
\end{eqnarray}
and
\begin{eqnarray}
\alpha=n+\frac{1}{2},\;\;\;\;\;
V^{(n)}_0=\frac{d^nV}{dr^n_*}\bigg|_{\;r_*=r_*(r_{p})} \nonumber.
\end{eqnarray}

\begin{table}[h]
\caption{The low overtones quasinormal frequencies of the massless
scalar field in the  black hole surrounded by quintessence for
fixed $l=5$ and $c=0.001$.}
\begin{tabular}[b]{c|c|c|c|c|c}
 \hline \hline
 $3\epsilon+1$ &$\omega\;(n=0)$& $\omega \;(n=1)$ &$\omega \;(n=2)$ &
 $\omega \;(n=3)$&$\omega \;(n=4)$  \\ \hline
   0&1.05797-0.09641i&1.04839-0.29037i&1.03048-0.48729i&1.00601-0.68789i&0.97651-0.89188i
 \\
-0.2&1.05757-0.09643i&1.04800-0.29043i&1.03010-0.48738i&1.00565-0.68801i&0.97617-0.89204i
 \\
-0.4&1.05707-0.09645i&1.04751-0.29050i&1.02962-0.48751i&1.00520-0.68818i&0.97575-0.89224i
 \\
-0.6&1.05645-0.09649i&1.04690-0.29061i&1.02904-0.48768i&1.00465-0.68842i&0.97524-0.89253i
\\
-0.8&1.05567-0.09654i&1.04614-0.29075i&1.02832-0.48792i&1.00398-0.68870i&0.97463-0.89291i
\\
-1.0&1.05470-0.09661i&1.04520-0.29096i&1.02743-0.48824i&1.00318-0.68916i&0.97392-0.89342i
\\
-1.2&1.05349-0.09670i&1.04404-0.29124i&1.02636-0.48869i&1.00222-0.68973ii&0.97310-0.89409i
\\
-1.4&1.05198-0.09684i&1.04260-0.29163i&1.02507-0.48930i&1.00111-0.69051i&0.97219-0.89497i
\\
-1.6&1.05009-0.09703i&1.04084-0.29218i&1.02352-0.49013i&0.99983-0.69153i&0.97119-0.89609i
\\
-1.8&1.04774-0.09730i&1.03868-0.29294i&1.02170-0.49126i&0.99841-0.69287i&0.97014-0.89748i
\\
-2.0&1.04481-0.09768i&1.03604-0.29400i&1.01956-0.49279i&0.99685-0.69459i&0.96912-0.89917i
\\
\hline \hline
\end{tabular}
 \end{table}
\begin{table}[h]
\caption{The low overtones quasinormal frequencies of the massless
scalar field in the Schwarzschild black hole for fixed $l=5$.}
\begin{tabular}[b]{c|c|c|c|c}
 \hline \hline
 $\omega\;(n=0)$& $\omega \;(n=1)$ &$\omega \;(n=2)$ &
 $\omega \;(n=3)$&$\omega \;(n=4)$  \\
 \hline
1.05957-0.09634i& 1.04997-0.29017i&1.03201-0.48696i&
1.00748-0.68744i&0.97789-0.89134i
 \\
\hline \hline
\end{tabular}
 \end{table}

Substituting the effective potential (\ref{v}) into the formula
above, we can obtain the quasinormal frequencies for the scalar
field in the background of the black hole surrounded by
quintessence. The lowly decaying modes frequencies for fixed $l=5$
and $c=0.001$ are list in the Table 1. From Fig.2, we also find
that for fixed $n$ and $l$ the absolute value of the imaginary
part decreases as the real part increases.  Moreover, Fig. 3 and 4
tell us that as the state parameter $\epsilon$ increases, the real
part increases and the absolute value of the imaginary part
decreases. It means that for larger $\epsilon$, the scalar field
decay more slowly. At last, we make a comparison between the
tables 1 and 2, and find that the real part in the Schwarzschild
spacetime is larger and the magnitude of imaginary parts is
smaller. In other words, due to presence of quintessence, the
oscillations damp more rapidly.

\begin{figure}[h]
\begin{center}
\includegraphics[width=6cm]{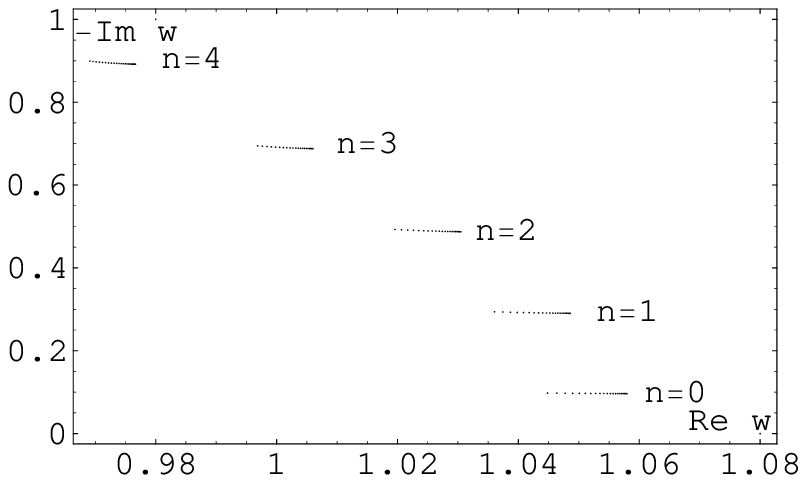}\;\;\;\;\;\includegraphics[width=6cm]{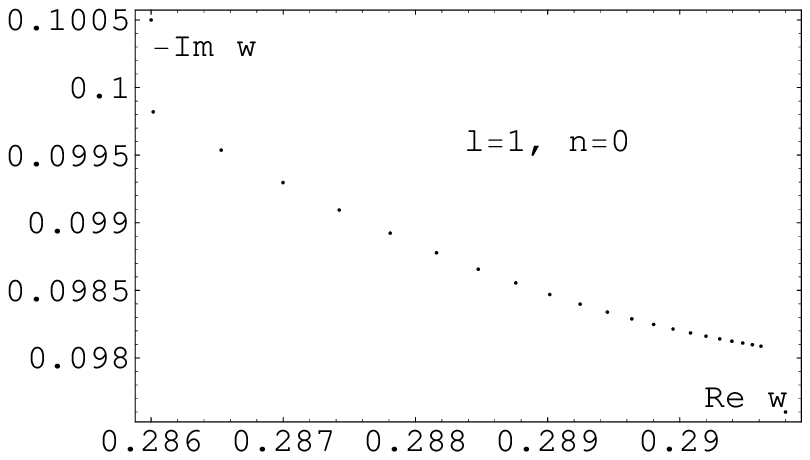}
\caption{The relationship between the real and imaginary parts of
quasinormal frequencies of the scalar field in the background of
the black hole surrounded by quintessence for fixed $c=0.001$. The
left is for $l=1$ and $n=0$, and the right is for $l=5$. }
 \end{center}
 \label{fig2}
 \end{figure}

\begin{figure}[h]
\begin{center}
\includegraphics[width=6cm]{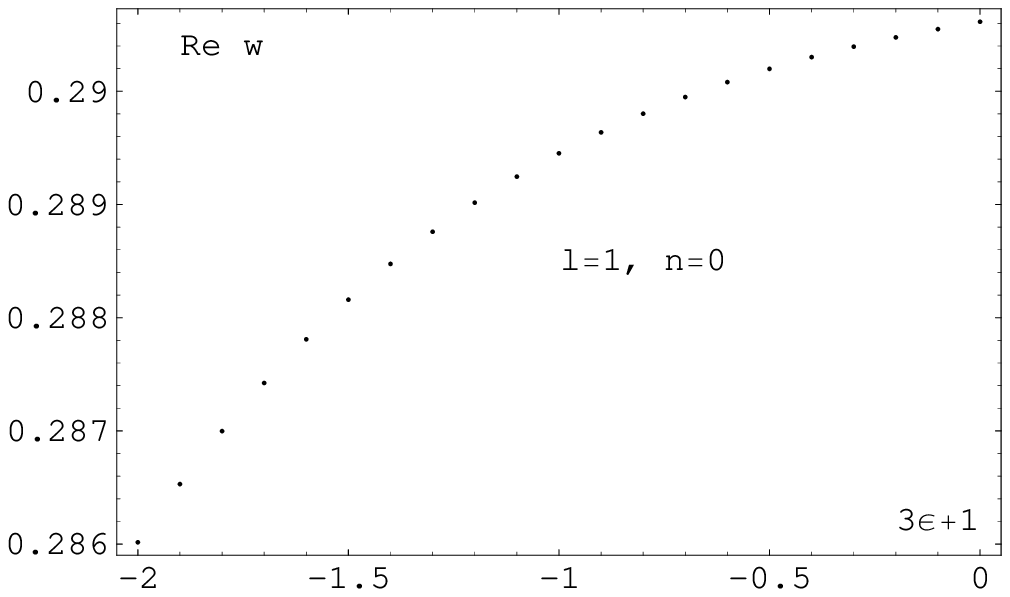}\;\;\;\includegraphics[width=6cm]{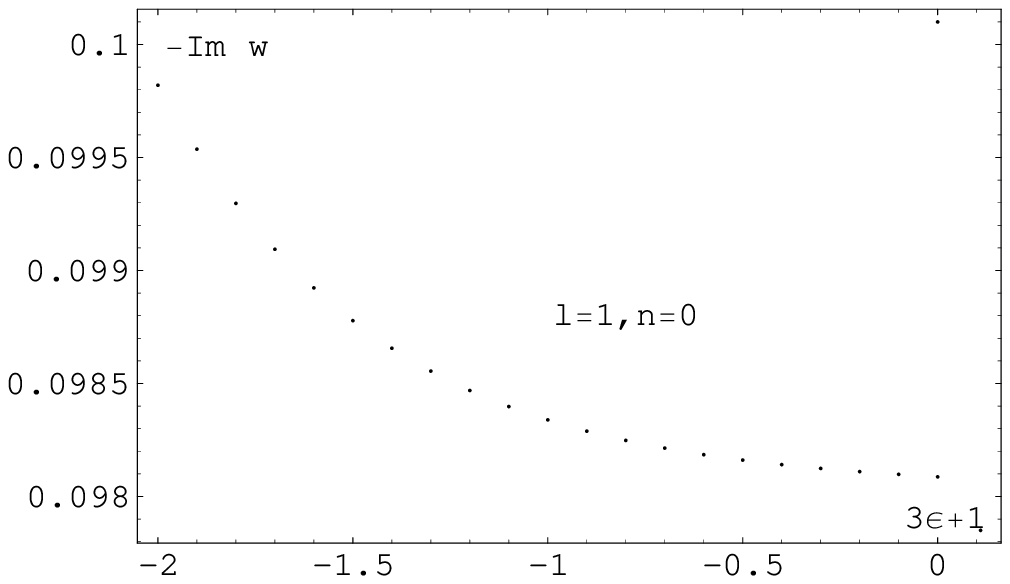}
\caption{Variation of the real (the left) and imaginary (the
right) parts of quasinormal frequencies of the scalar field with
the state parameter $\epsilon$ for $l=1$ and $n=0$.}
\end{center}
\label{fig3}
\end{figure}

\begin{figure}[h]
\begin{center}
\includegraphics[width=6cm]{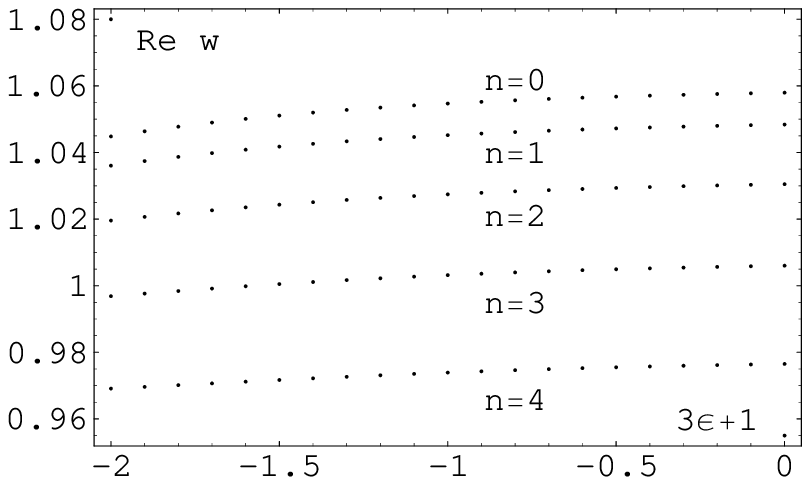}\;\;\;\includegraphics[width=6cm]{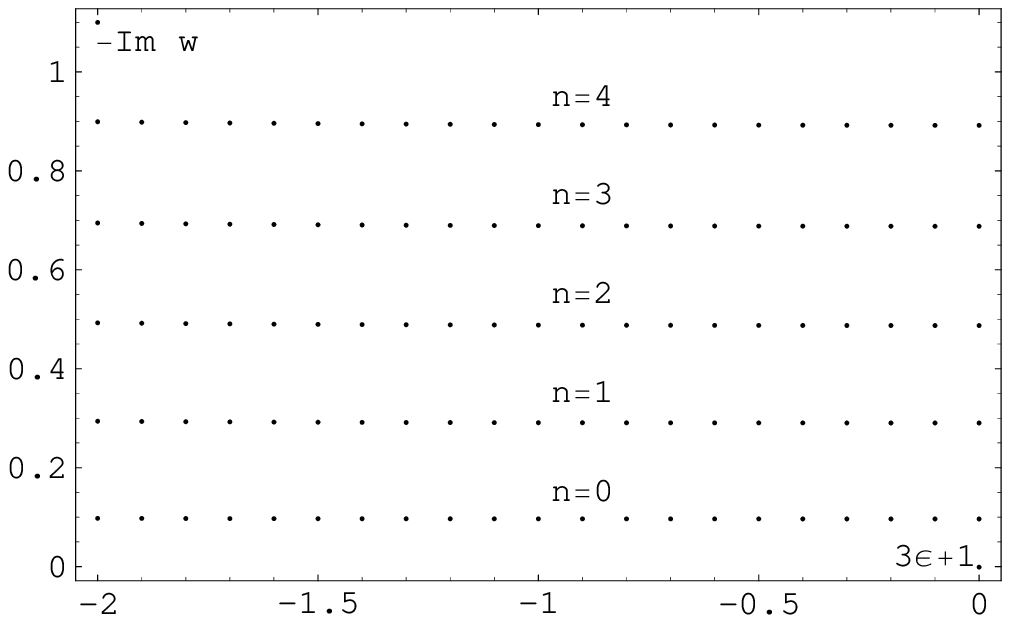}
\caption{Variation of the real (the left) and imaginary (the
right) parts of quasinormal frequencies of the scalar field with
the state parameter $\epsilon$ for $l=5$.}
 \end{center}
 \label{fig4}
 \end{figure}

 \vspace*{0.2cm}
\section{summary}

Using the third-order WKB approximation, we evaluated the
quasinormal modes of the scalar field in the black hole spacetime
surrounded by quintessence. Our results show that due to the
presence of quintessence, the scalar field damps more rapidly.
Moreover, we also note that the state parameter $\epsilon$ play an
important role for the quasinormal frequencies. As the state
parameter $\epsilon$ increases the real part increases and the
absolute value of the imaginary part decreases. This means that
the scalar field decays more slowly in the larger $\epsilon$
quintessence.

\begin{acknowledgments}
This work was supported by the National Natural Science Foundation
of China under Grant No.10275024, the FANEDD under Grant No.200317
and the Hunan Provincial Natural Science Foundation of China under
Grant No.04JJ3019, and the Hunan Normal University Natural Science
Foundation under Grant No.22040639.
\end{acknowledgments}


\newpage
\begin{thebibliography}{99}

\bibitem{AAS}A. G. Riess et al., Astron. J. {\bf 116}, 1009 (1998); S. Perl- mutter et
al., Astrophys. J. {\bf 517}, 565 (1999); A. G. Riess et al.,
Astrophys. J. {\bf 607}, 665 (2004); R. A. Knop et al., Astrophys.
J. {\bf 598}, 102 (2003).

\bibitem{PAS} P. de Bernardis et al., Nature {\bf 404},  955 (2000); A. D. Miller et
al. Astrophys. J. Lett. {\bf 524},  L1 (1999); S. Hanany et al.,
Astrophys. J. Lett. {\bf 545}, L5 (2000);  N. W. Halverson et al.,
Astrophys. J. {\bf 568}, 38 (2002); B. S. Mason et al., Astro-
phys. J. {\bf 591}, 540 (2003); D. N. Spergel et al., Astrophys.
J. Suppl. {\bf 148}, 175 (2003); L. Page et al., Astrophys. J.
Suppl. {\bf 148}, 233 (2003).

\bibitem{RMT} R. Scranton et al., astro-ph/0307335; M. Tegmark et al., Phys.
Rev. D {\bf 69}, 103501 (2004).

\bibitem{PBTT} P. J. E. Peebles, B. Ratra, Rev. Mod. Phys. {\bf 75}, 559 (2003); T.
Padmanabhan, Phys. Rept. {\bf 380}, 235 (2003); T. Padmanabhan,
Curr.Sci. {\bf 88}, 1057 (2005).

\bibitem{Swei} S. Weinberg, Rev. Mod. Phys. {\bf 61}, 1 (1989).

\bibitem{BRa} B. Ratra and  P. J. E. Peebles, Phys. Rev. D {\bf 37}, 3406 (1988);
P. J. E. Peebles and B. Ratra, Astrophys. J. {\bf 325}, L17
(1988); C. Wetterich, Nucl. Phys. B {\bf 302}, 668 (1988); R. R.
Caldwell, R. Dave, and P. J. Steinhardt, Phys. Rev. Lett. {\bf
80}, 1582 (1998); I. Zlatev, L. Wang and P. J. Steinhardt, Phys
Rev. Lett. {\bf 82}, 896 (1999); M. Doran and J. Jaeckel, Phys.
Rev. D {\bf 66}, 043519 (2002).

\bibitem{CAR} C. A. Picon,
T. Damour and V. Mukhanov, Phys. Lett. B {\bf 458}, 209 (1999); T.
Chiba, T. Okabe and M. Yamaguchi, Phys. Rev. D {\bf 62}, 023511
(2000).

\bibitem{RRCa} R. R. Caldwell, Phys. Lett. B {\bf 545}, 23 (2002).

\bibitem{BMcI} B. McInnes, JHEP {\bf 0208}, 029 (2002); S. Nojiri and S. D. Odintsov,
Phys. Lett. B {\bf 562}, 147 (2003); S. Nojiri and S. D. Odintsov,
Phys. Lett B {\bf 565}, 1 (2003); P. Singh, M. Sami, and N.
Dadhich, Phys. Rev. D {\bf 68}, 023522 (2003); G. W. Gibbons,
hep-th/0302199; L. P. Chimento and R. Lazkoz, Phys. Rev. Lett.
{\bf 91}, 211301 (2003);

\bibitem{Hawking}S. W. Hawking, Nature {\bf 248}, 30 (1974);
Commun. Math. Phys. {\bf 43}, 199 (1975).

\bibitem{JDB} J. D. Bekenstein, Phys.
Rev. D {\bf 7}, 2333-2346 (1973).

\bibitem{GHa}G. W. Gibbons and S. W. Hawking, Phys. Rev. D {\bf 15},
2738-2751 (1977).

\bibitem{Scr} S. Carlip, Phys. Rev. D {\bf 51}, 632-637 (1995); S. Carlip
 Phys. Rev. D {\bf 55}, 878-882 (1997).

\bibitem{Hooft}G.'t Hooft,
Nuclear Phys. B {\bf 256}, 727-745 (1985).

\bibitem{SHa} S. Hawking, Phys. Rev. D {\bf 14}, 2460 (1976).

\bibitem{Hooft1} G.'t Hooft, Nucl. Phys. B {\bf 335}, 138 (1990).

\bibitem{UHD}U. H. Danielsson, Phys. Rev. D {\bf 48}, 4779-4784
(1993).

\bibitem{SHod1} S. Hod, Phys. Lett. A {\bf  299}, 144-148 (2002).


\bibitem{JDB1} J. D. Bekenstein, Phys. Rev. D {\bf 23}, 287 (1981).

\bibitem{RBOu} R. Bousso,  JHEP {\bf 9906}, 028 (1999); JHEP {\bf 9907}, 004
(1999);  Rev. Mod. Phys. {\bf 74}, 825 (2002)

\bibitem{LSU} L. Susskind, J. Math. Phys {\bf 36}, 6377 (1995).

\bibitem{Chandrasekhar}S. Chandrasekhar and S. L. Detweiler,
Proc. R. Soc. London  A {\bf 344}, 441 (1975).

\bibitem{Regge} T. Regge and  J. A. Wheeler, Phys. Rev. {\bf 108}
1063 (1957).

\bibitem{Witten} E. Witten,  Adv. Theor. Math. Phys.
  {\bf  2},  253 (1998).

\bibitem{Maldacena} J. Maldacena, Adv. Theor. Math. Phys.
  {\bf  2},  231 (1998).

\bibitem{Kalyana} S. Kalyana Rama and Sathiapalan, Mod Phys. Lett.
 A {\bf  14},  2635 (1999).

\bibitem{Hod98} S. Hod,
 Phy. Rev. Lett. {\bf 81}, 4293 (1998).

\bibitem{Dreyer03}O. Dreyer,
 Phy. Rev. Lett. {\bf 90}, 081301 (2003).

\bibitem{Cardoso} V. Cardoso and J. P. S. Lemos, Phys. Rev. D {\bf 63}
124015 (2001).

\bibitem{Konoplya} R. A. Konoplya, Phys. Rev. D {\bf 66}  084007 (2002).

\bibitem{Starinets} A. O. Starinets, Phys. Rev. D {\bf 66}  124013 (2002).

\bibitem{Setare} M. R. Setare, Class. Quant. Grav. {\bf 21}  1453
(2003); Phys. Rev. D {\bf 69}  044016 (2004).

\bibitem{Natario} J. Natario and R. Schiappa, hep-th/0411267.

\bibitem{Leaver} E. W. Leaver, Pro. R. Soc. Lond. {\bf A} 402, 285
(1985);  E. W. Leaver, Phys. Rev. D 34, 384 (1986).

\bibitem{Cho} H. T. Cho, Phys. Rev. D {\bf 68}, 024003 (2003).

\bibitem{Zhidenko}A. Zhidenko, Class. Quant. Grav. {\bf 21} 273
(2004).

\bibitem{Moss} I. G. Moss and J. P. Norman, Class. Quantum Grav. {\bf 19}, 2323
(2002); F. Mellor and I. G. Moss, Phys. Rev. D {\bf 41}, 403
(1990).


\bibitem{Jing1} J. L. Jing,   Phys. Rev. D {\bf 69}, 084009 (2004).


\bibitem{Branco}K. H. C. Castello-Branco, R. A. Konoplya, and A.
Zhidenko,  Phys. Rev. D {\bf 71},  047502 (2005).

\bibitem{Jing2} J. L. Jing,  Phys. Rev. D {\bf 70}, 065004
(2004); Phys. Rev. D {\bf 71},  124011 (2005);
 Phys. Rev. D {\bf 71}, 124006 (2005).


\bibitem{VVK} V. V. Kiselev, Class. Quant. Grav. {\bf 20},
1187-1198 (2003).

\bibitem{BC}B. F. Schutz and C. M. Will,
Astrophys. J. Lett. Ed. {\bf 291}, L33 (1985).

\bibitem{SC}S. Iyer and C. M. Will,
Phys. Rev. D  {\bf 35}, 3621 (1987).

\bibitem{SI}S. Iyer, Phys. Rev. D {\bf 35}, 3632 (1987).

\end{thebibliography}
\end{document}